\documentclass[aps,prd,floatfix,superscriptaddress,nofootinbib,12pt]{revtex4-1}

\usepackage[colorlinks=true, linkcolor=blue, urlcolor=blue, citecolor=green,          
            bookmarks=true, bookmarksnumbered=true, breaklinks=true, pdfpagemode=Fullscreen,pdfstartview=FitBH]{hyperref}
\usepackage{epsfig,latexsym,cancel,amssymb,amsmath,verbatim,mathrsfs}
\usepackage{subcaption}
\usepackage{graphicx}
\usepackage{bm}
\usepackage{xcolor}
\usepackage{multirow}
\usepackage{slashed}
\allowdisplaybreaks
\usepackage{siunitx}
\usepackage{diagbox}
\usepackage{tablefootnote}
\usepackage{tabularray}

\usepackage{tikz}
\usetikzlibrary{arrows,shapes}
\usetikzlibrary{trees}
\usetikzlibrary{matrix,arrows} 				
\usetikzlibrary{positioning}				
\usetikzlibrary{calc,through}				
\usetikzlibrary{decorations.pathreplacing}  
\usepackage{pgffor}							
\usetikzlibrary{decorations.pathmorphing}	
\usetikzlibrary{decorations.markings}
\tikzset{
    vector/.style={decorate, decoration={snake}, draw},
	provector/.style={decorate, decoration={snake,amplitude=2.5pt}, draw},
	antivector/.style={decorate, decoration={snake,amplitude=-2.5pt}, draw},
    fermion/.style={draw=black, postaction={decorate},
        decoration={markings,mark=at position .55 with {\arrow[draw=black,line width=2pt]{stealth}}}},
    fermionbar/.style={draw=black, postaction={decorate},
        decoration={markings,mark=at position .55 with {\arrow[draw=black,line width=2pt]{stealth[reversed]}}}},
    fermionnoarrow/.style={draw=black},
    gluon/.style={decorate, draw=black,
        decoration={coil,amplitude=4pt, segment length=5pt}},
    scalar/.style={dashed,draw=black, postaction={decorate},
        decoration={markings,mark=at position .55 with {\arrow[draw=black,line width=2pt]{stealth}}}},
    scalarbar/.style={dashed,draw=black, postaction={decorate},
        decoration={markings,mark=at position .55 with {\arrow[draw=black,line width=2pt]{stealth[reversed]}}}},
    scalarnoarrow/.style={dashed,draw=black},
    electron/.style={draw=black, postaction={decorate},
        decoration={markings,mark=at position .55 with {\arrow[draw=black,line width=2pt]{stealth}}}},
	bigvector/.style={decorate, decoration={snake,amplitude=4pt}, draw},
}
\tikzstyle{block} = [draw, rectangle, 
    minimum height=3em, minimum width=6em]

\newcommand{\calO}{{\cal O}}
\newcommand{\calB}{{\cal B}}

\begin{document}
\baselineskip=17pt \parskip=5pt

\hspace*{\fill}

\title{ Search for Light Dark Matter in Rare Meson Decays}

\author{Ze-Kun Liu}
\email{liuzekun@htu.edu.cn}
\affiliation{Institute of Particle and Nuclear Physics,
                 Henan Normal University, Xinxiang 453007, China}

\author{Ying Li}
\email{liying20239@stu.htu.edu.cn}
\affiliation{Institute of Particle and Nuclear Physics,
                 Henan Normal University, Xinxiang 453007, China}
                 
\author{Biao-Feng Hou}
\email{resonhou@zknu.edu.cn}
\affiliation{Institute of Particle and Nuclear Physics,
                 Henan Normal University, Xinxiang 453007, China}       
                 
\author{Qin Chang}
\email{qinchang@htu.edu.cn}
\affiliation{Institute of Particle and Nuclear Physics,
                 Henan Normal University, Xinxiang 453007, China}

\begin{abstract}
Current dark matter direct detection experiments have low sensitivity to sub-GeV dark matter. In this work, we demonstrate that rare $B$ and $K$ meson decays with missing energy in the final state can serve as efficient probes in this mass range. We analyze a generic $Z^{\prime}$ portal dark matter model and derive upper limits on its parameters from experimental bounds on the rare $B$ and $K$ meson decays. Our results show that such meson decay processes provide complementary constraints to current direct detection experiments for sub-GeV dark matter, particularly for interaction forms mediated by dark matter momentum-dependent operators.

\end{abstract}

\maketitle

\newpage

\section{Introduction}
Abundant astronomical and cosmological evidence indicates that about a quarter of the matter in the Universe is composed of dark matter (DM)~\cite{ParticleDataGroup:2024cfk}. However, the nature of DM remains unknown as the Standard Model (SM) of particle physics fails to provide a viable DM candidate. The most widely studied DM candidates are the so-called weakly interacting massive particles (WIMPs), which could naturally provide the right amount of DM relic density, a coincidence known as the WIMP miracle. Most current DM direct detection experiments target the search for WIMPs, with mass range roughly on the order of GeV to TeV. DM direct detection experiments mostly aim to observe elastic or inelastic scattering of DM particles with atomic nuclei, such as liquid xenon detectors XENONnT~\cite{XENON:2023cxc}, LUX-ZEPLIN~\cite{LZ:2022lsv}, and PandaX-4T~\cite{PandaX:2022xas}. While no confirmed DM signal has been observed, these experiments have placed stringent limits on the DM–nucleon scattering cross section. However, their sensitivity drops sharply for DM masses below the GeV scale, as the nuclear recoil signals from its elastic scattering with nuclei typically lie below the detection energy threshold and are overwhelmed by background noise. Searching for light DM in heavy meson decays can serve as a complement to DM direct detection, and its sensitivity in some models may be higher than that of direct detection experiments.

Semileptonic flavor-changing neutral current (FCNC) decays are highly suppressed in the SM due to their loop origin and the Glashow-Iliopoulos–Maiani (GIM) mechanism, making them exceptionally sensitive to new physics (NP) effects. In particular, decays with a neutrino pair in the final state provide a distinct probe of NP compared to those with charged leptons. For semileptonic decays to charged leptons, NP effects are primarily probed through modifications of the relevant Wilson coefficients, mediated by virtual NP particles. In contrast, for decays to a neutrino pair, NP particles themselves can exist as missing energy hiding in the final state, since the neutrinos are not detected in experiments. In addition to the SM neutrinos, any electrically neutral, undetectable particle can manifest as missing energy in such processes. Notably, the Belle II experiment has recently reported the first evidence of $B^+\to K^+\nu\bar{\nu}$ with the branching ratio of $\text{Br}(B^+\to K^+\nu\bar{\nu}) =(23\pm5^{+5}_{-4})\times10^{-6}$~\cite{Belle-II:2023esi}, which is about $2.7 \sigma$ above the SM prediction $(4.16 \pm 0.57) \times10^{-6}$~\cite{Hou:2024vyw,Altmannshofer:2009ma,Buras:2014fpa, Becirevic:2023aov}. Consequently, it has led to numerous studies exploring explanations beyond the SM, with possible excess sources including sterile neutrinos~\cite{Felkl:2023ayn,Buras:2024ewl,Datta:2023iln,Rosauro-Alcaraz:2024mvx}, axion-like particle~\cite{Li:2024thq,Li:2025ski,Calibbi:2025rpx,Altmannshofer:2024kxb}, dark matter~\cite{He:2022ljo, He:2023bnk,He:2024iju,Aliev:2025hyp,He:2025jfc,Altmannshofer:2023hkn, Hou:2024vyw, Fridell:2023ssf,Gabrielli:2024wys,Berezhnoy:2025osn,MartinCamalich:2025srw,Abdughani:2023dlr,Wang:2023trd} or modifications to the relevant Wilson coefficients from other NP~\cite{Sumensari:2024sji, Chen:2024jlj, Allwicher:2023xba,Athron:2023hmz,Bause:2023mfe,Chen:2024cll}. The primary aim of this work is to derive constraints on a generic $Z^{\prime}$ portal dark matter model from rare meson decays with a neutrino pair in the final state and explore the complementarity between such decays and direct detection experiments for different DM–quark operator structures in the model.

This paper is organized as follows. In section~\ref{sec:Theoretical Framework}, we establish the theoretical framework of a generic $Z^{\prime}$ portal model with all possible renormalizable couplings to quarks and DM. In section \ref{sec:BKdecay}, we present a systematic analysis of the rare meson decay processes, such as $B \to P(V)\chi\bar{\chi}$ and $K \to \pi\chi\bar{\chi}$. Using the latest experimental data, we derive joint constraints on the coupling strength and mass of light DM. In section~\ref{sec:complementarity}, we compute the corresponding constraints on the cross section between light DM and nucleons, using the upper limits on the $Z^{\prime}$ portal model parameters derived from rare meson decays. By comparing these constraints with those from direct detection experiments, we assess the complementarity between these two approaches. Summary and conclusions are given in the last section~\ref{sec:conclusion}.

\section{Theoretical Framework}
\label{sec:Theoretical Framework}

The WIMP miracle strongly suggests that the interaction strength between DM and SM particles is most likely at the electroweak scale. However, DM that interacts with the SM purely through electroweak interactions is strongly constrained~\cite{DeSimone:2014qkh}. Therefore, extending the SM with a new $U(1)$ gauge symmetry to describe their interactions is one of the simplest and best motivated approaches. In this paper, we study a generic $Z^{\prime}$ portal model, in which a heavy electrically neutral $Z^{\prime}$ serves as the portal for DM. We consider a Dirac fermion $\chi$ as the DM particle that interacts with the $Z^{\prime}$. The general model-independent Lagrangian can be written as 
\begin{equation}
\mathcal{L}_{\mathrm{int}}=
  Z_{\mu}^{\prime}\bar{f}\gamma^{\mu}
    \left(g_{f}^V + g_{f}^A \,\gamma_{5}\right) f 
  + Z_{\mu}^{\prime}\bar{\chi}\gamma^{\mu}
    \left(g_{\chi}^V + g_{\chi}^A \, \gamma_{5}\right) \chi\,,
    \label{eq:Lagrangian}
\end{equation}
where $f$ denotes all quarks in the SM. Note that the $Z^{\prime}$ boson can also couple to leptons in the SM, corresponding to a leptophilic dark matter model. The collider experiments impose stringent constraints on the mass of the $Z^{\prime}$ boson if it couples simultaneously to both quarks and leptons~\cite{Alves:2015pea}. In this work, we mainly focus on the complementarity of rare meson decays to the parameter space of DM direct detection experiments. Therefore, we consider the scenario where dark matter interacts directly only with quarks via the $Z^{\prime}$ boson. This setup can be realized by assuming that the $U(1)$ charges of leptons are zero.
 
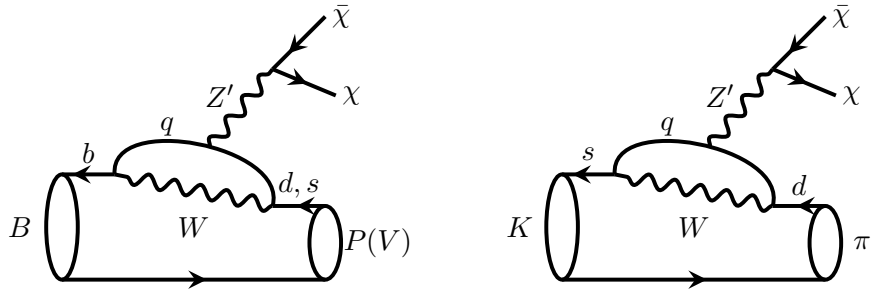
\begin{figure}[htbp]
\centering
\begin{tikzpicture}[line width=1.5pt, scale=0.7]
\begin{scope}
    \draw (0,0) ellipse [x radius=0.3, y radius=1];
	\draw[fermionbar] (0,1)--(1,1);
    \draw[fermion] (0,-1)--(5,-1);
    \draw (5,-0.3) ellipse [x radius=0.3, y radius=0.7];
    \draw[fermion] (5,0.4)--(4,0.4);
    \draw[vector] (4,0.4)--(1,1);
     \draw [-, black] (4,0.4) to [in=95, out=75] (1,1);
    \draw[vector] (2.8,1.5)--(4,3);
	\draw[fermionbar] (4,3)--(5,4);
    \draw[fermion] (4,3)--(5.2,2.5);
    \node at (-0.8,0) {$B$};
    \node at (6,-0.3) {$P(V)$};
    \node at (2.5,0) {$W$};
    \node at (4.5,0.8) {$d,s$};
    \node at (0.5,1.4) {$b$};
    \node at (2,1.9) {$q$};
     \node at (3,2.5) {$Z^{\prime}$};
    \node at (5.3,4) {$\bar{\chi}$};
    \node at (5.5,2.5) {$\chi$};
\end{scope}
\end{tikzpicture}
\hspace{1cm}
\begin{tikzpicture}[line width=1.5pt, scale=0.7]
\begin{scope}
    \draw (0,0) ellipse [x radius=0.3, y radius=1];
	\draw[fermionbar] (0,1)--(1,1);
    \draw[fermion] (0,-1)--(5,-1);
    \draw (5,-0.3) ellipse [x radius=0.3, y radius=0.7];
    \draw[fermion] (5,0.4)--(4,0.4);
    \draw[vector] (4,0.4)--(1,1);
     \draw [-, black] (4,0.4) to [in=95, out=75] (1,1);
    \draw[vector] (2.8,1.5)--(4,3);
	\draw[fermionbar] (4,3)--(5,4);
    \draw[fermion] (4,3)--(5.2,2.5);
    \node at (-0.8,0) {$K$};
    \node at (5.7,-0.3) {$\pi$};
    \node at (2.5,0) {$W$};
    \node at (4.5,0.8) {$d$};
    \node at (0.5,1.4) {$s$};
    \node at (2,1.9) {$q$};
    \node at (3,2.5) {$Z^{\prime}$};
    \node at (5.3,4) {$\bar{\chi}$};
    \node at (5.5,2.5) {$\chi$};
\end{scope}
\end{tikzpicture}
 \caption{Leading-order Feynman diagrams for the $B$ and $K$ meson decays into a pair of DM, with $q=u,c,t$.}
\label{fig:feynman}
\end{figure}

In this scenario, the $B$ and $K$ meson decays with a pair of DM in the final states correspond to the processes shown in Figure~\ref{fig:feynman}. At the quark level, they arise from the $b(s) \to s(d)$ transitions through penguin diagrams. 
This is distinct from the case where non-universal $Z^{\prime}$-mediated quark flavor-changing transitions can occur at the tree level. In our framework, flavor-changing processes are generated at the one-loop level and another key feature is the generation-universal coupling of the $Z^{\prime}$ boson. Thus, constraints on the couplings $g_f^{V(A)}$ and $g_{\chi}^{V(A)}$ from rare meson decays must account for the total contributions from all three-generation fermions.

\section{$B$ And $K$ Meson Decays}
\label{sec:BKdecay}
In this section, we derive bounds on the parameters of the $Z^{\prime}$ portal model from the rare $B$ and $K$ meson decays. For convenience, we will use the dark low-energy effective field theory (DLEFT)~\cite{Aebischer:2022wnl,Song:2023jqm} to describe these meson FCNC decays, and then match the simplified model described by Eq.~\eqref{eq:Lagrangian} onto the DLEFT at the meson decay scales. Next, we will first briefly introduce the DLEFT and give the matching conditions. The analytic and numerical analysis of the differential branching ratio of $B$ and $K$ meson decays is given in the end of this section.

\subsection{Effective Hamiltonian}
\label{sec:DLEFT}
As we know, the low-energy effective field theory(LEFT)~\cite{Jenkins:2017jig,Liang:2023yta} is suitable for describing meson decays. The DLEFT is built on LEFT by adding new effective operators to describe the interactions between DM and SM sectors~\cite{Aebischer:2022wnl,Song:2023jqm}. Since new physics effects arise from $Z^\prime$ in this work, it will only induce vector and axial-vector quark currents at low-energy scales. Restricting to dimension-6 operators, the effective Hamiltonian governing the $d_i\to d_j\chi\bar{\chi}$ transitions takes the form
\begin{align}
   \mathcal{H}_\mathrm{eff} =  C_{d\chi}^{V V} \calO_{d\chi}^{V V} 
     + C_{d\chi}^{V A} \calO_{d\chi}^{V A} 
     + C_{d\chi}^{A V} \calO_{d\chi}^{A V}
     + C_{d\chi}^{A A} \calO_{d\chi}^{A A}\,,
  \label{eq:DLEFT Hamiltonian}
\end{align}
in which $C_{d\chi}$s are Wilson coefficients and the operators are written by
\begin{align}
    \begin{aligned}
    \mathcal{O}_{d\chi}^{V V} &= 
        (\bar{d}_i\gamma^\mu d_j)(\bar{\chi}\gamma_\mu\chi)\,, &
    \mathcal{O}_{d\chi}^{V A} &= 
        (\bar{d}_i\gamma^\mu d_j)(\bar{\chi}\gamma_\mu\gamma_5\chi)\,,\\
    \mathcal{O}_{d\chi}^{A V} &= 
        (\bar{d}_i \gamma^\mu\gamma_5 d_j)(\bar{\chi}\gamma_\mu\chi)\,,&
    \mathcal{O}_{d\chi}^{A A} &=
        (\bar{d}_i \gamma^\mu\gamma_5 d_j)(\bar{\chi}\gamma_\mu\gamma_5\chi)\,.
    \end{aligned}
     \label{eq:opertor}
\end{align}
Here, $d$ denotes the down-type quarks in the mass basis, with $i$, $j$ the flavor indices. $\chi$ denotes the fermion-type DM candidate.

Given the full theory at the new physics scale (see Eq.~\eqref{eq:Lagrangian} for details), we can match it onto the effective Hamiltonian at the $B$ or $K$ meson scale. Then, the Wilson coefficients in Eq.~\eqref{eq:DLEFT Hamiltonian} can be written in terms of the mass and the couplings of $Z^\prime$. The detailed results are presented as
\begin{align}
  \begin{aligned}
  C_{d\chi}^{V V} &= \frac{g^2_W g_{f}^{V}g_{\chi}^V}{256\pi^{2}m_{Z^\prime}^2}\lambda_q^{i,j} F_1(x_q)\,,&
  C_{d\chi}^{V A} &= \frac{g^2_W g_{f}^{V}g_{\chi}^A}{256\pi^{2}m_{Z^\prime}^2}\lambda_q^{i,j} F_1(x_q)\,,\\
  C_{d\chi}^{A V} &= \frac{g^2_W g_{f}^{A}g_{\chi}^V}{256\pi^{2}m_{Z^\prime}^2}\lambda_q^{i,j} F_2(x_q)\,,&
  C_{d\chi}^{A A} &= \frac{g^2_W g_{f}^{A}g_{\chi}^A}{256\pi^{2}m_{Z^\prime}^2}\lambda_q^{i,j} F_2(x_q)\,,
  \end{aligned}
  \label{eq:matching}
 \end{align}
in which $g^2_W=4\sqrt{2}m_W^2G_F$ with $m_W$ the $W$ boson mass and $G_F$ the Fermi constant. The CKM factor $\lambda_q^{i,j}=V_{qi}^{*}V_{qj}$ and $x_q = m_q^2/m_W^2$ with $q=u,\ c,\ t$. Note that the interaction strength between dark matter and quarks can be described by an effective point interaction when the $Z^{\prime}$ mass is heavy. Therefore, the complementarity between meson decays and direct dark matter detection is independent of the value of $m_{Z^\prime}$ and we take a typical mass $m_{Z^\prime}=100$ GeV in this work. The loop functions are given by
\begin{align}
\begin{aligned}
   F_1(x)&=5x-\frac{6}{x-1}-\frac{6x^{2}(x-2)}{(x-1)^{2}}\ln x +6x\ln\left[\frac{\mu^{2}}{m_\mathrm{W}^{2}}\right]\,,\\
   F_2(x)&=-\frac{3\left(x^2+3x-2\right)}{x-1}+\frac{2x\left(5x^2-10x+8\right)}{(x-1)^{2}}\ln x - 10x \ln\left[\frac{\mu^{2}}{m_\mathrm{W}^{2}}\right]\,.
\end{aligned}
\end{align}
A renormalization scale $\mu$ is used to regularize the UV-divergent loop contribution. We employ $\mu_B\approx4.8$ GeV for $B$ meson decays and $\mu_K\approx0.49$ GeV for $K$ meson decays.

\begin{table}
\centering
\resizebox{\linewidth}{!}{
\renewcommand{\arraystretch}{1.1}
\begin{tabular}{ c  c   c  c  }
\hline
&  Observable \quad\quad  
&  SM \quad\quad
&  Exp \quad\quad 
\\\hline 
  \multirow{4}{*}{\quad\quad $b\rightarrow s$ \quad\quad}
& $\calB(B^+ \to K^+ \nu\bar{\nu}) $\quad\quad
& $(4.16 \pm 0.57) \times10^{-6}$\cite{Hou:2024vyw,Altmannshofer:2009ma,Buras:2014fpa, Becirevic:2023aov}\quad\quad
&$(23\pm5^{+5}_{-4}) \times10^{-6}$\cite{Belle-II:2023esi} \\
  \cline{2-4}
&$\calB(B^{0}\rightarrow K^{0}\nu\bar{\nu})$\quad\quad
& $(4.4\pm 0.7)\times10^{-6}$\quad\quad
&$<26\times10^{-6}$\cite{ParticleDataGroup:2024cfk,Belle:2017oht} \\
  \cline{2-4}
&$\calB(B^{+}\rightarrow K^{*+}\nu\bar{\nu}) $\quad\quad
&$(9.70\pm 0.94)\times10^{-6}$\quad\quad
&$<40\times10^{-6}$\cite{ParticleDataGroup:2024cfk,Belle:2013tnz}\\
   \cline{2-4}
&$\calB(B^{0}\to K^{*0}\nu\bar{\nu})$ \quad\quad
&$(9.00\pm 0.87)\times10^{-6}$\quad\quad
&$<18\times10^{-6}$\cite{ParticleDataGroup:2024cfk,Belle:2017oht}
\\\hline 
  \multirow{4}{*}{\quad\quad $b\rightarrow d$ \quad\quad}
&$\calB(B^{+} \rightarrow\pi^{+}\nu\bar{\nu}) $\quad\quad
&$(2.39^{+0.30}_{-0.28})\times10^{-7}$\cite{Hambrock:2015wka}\quad\quad
&$<140\times10^{-7}$\cite{ParticleDataGroup:2024cfk,Belle:2017oht}\\
 \cline{2-4}
&$\calB(B^{0} \rightarrow\pi^{0}\nu\bar{\nu})$\quad\quad
&$(1.2^{+0.15}_{-0.14})\times10^{-7}$\cite{Hambrock:2015wka}\quad\quad
&$<900\times10^{-8}$\cite{ParticleDataGroup:2024cfk,Belle:2017oht}\\
 \cline{2-4}
&$\calB(B^{+} \to\rho^{+}\nu\bar{\nu}) $\quad\quad
&$(4.06\pm 0.79)\times10^{-7}$\quad\quad
&$<300\times10^{-7}$\cite{ParticleDataGroup:2024cfk,Belle:2017oht}\\
 \cline{2-4}
&$\calB(B^{0} \to\rho^{0}\nu\bar{\nu}) $\quad\quad
&$(1.89\pm 0.36)\times10^{-7}$\quad\quad
&$<400\times10^{-7}$\cite{ParticleDataGroup:2024cfk,Belle:2017oht}
\\\hline 
  \multirow{1}{*}{\quad\quad $s\rightarrow d$ \quad\quad}
&$\calB(K^{+}\to\pi^{+}\nu\bar{\nu})$\quad\quad
&$(8.40\pm 1.00)\times10^{-11}$\quad\quad
&$(13.0_{-3.0}^{+3.3})\times10^{-11}$\cite{NA62:2024pjp}
\\\hline 
\end{tabular}
}
\caption{Summary of SM predictions and experimental measurements for $B$ and $K$ meson decays into a pair of neutrinos in the final state, where SM predictions lacking a reference are computed using the \texttt{flavio} package~\cite{Straub:2018kue}.}
\label{tab:Margin_settings}
\end{table}

\subsection{$B$ Meson Decay}
\label{sec:Bdecay}
In this subsection, we will analyze the differential decay widths of rare $B$ meson decays, as well as their constraints on the parameter space of the $Z^\prime$ portal model. The SM predictions and experimental measurements for the relevant meson decays are summarized in Table~\ref{tab:Margin_settings}. Experimentally, neutrino pairs in the final states of these processes appear as missing energy. At quark level, transitions induced by the $b\to s$ or $b\to d$ decay with invisible final states could contribute to these processes, allowing these rare decays to constrain the couplings between light DM and quarks.

In order to derive the decay rates for various processes using the DLEFT introduced in the previous subsection, an essential step is the determination of the relevant hadronic matrix elements. For $B\to P$ and $B\to V$ transitions, where $P$ denotes pseudoscalar meson (e.g., $\pi$ or $K$) and $V$ vector meson (e.g., $\rho$ or $K^*$), non-vanishing hadronic matrix elements from vector and axial-vector quark currents can be expressed in terms of the relevant form factors
\begin{align}
\begin{aligned}
    \langle P(p^{\prime}) | {\bar d_i} \gamma^\mu b | B(p) \rangle  &= 
    \left[ \left( p+p^{\prime} \right)^\mu - 
        \frac{m_B^2-m_P^2}{q^2} q^\mu \right] f_{+}(q^2) 
    + \frac{m_B^2-m_P^2}{q^2} q^\mu f_{0}(q^2)\,,\\
    \langle V(p^{\prime}) | {\bar d_i} \gamma^\mu b | B(p) \rangle  
        &= \epsilon^{\mu \nu \rho \sigma} \epsilon_{V,\nu}^* p_\rho p^{\prime}_{\sigma}, \frac{2}{m_B+m_V}V_0(q^2)\,,\\
    \langle V(p^{\prime})|\bar{d_i}\gamma^{\mu}\gamma_{5}b|B(p)\rangle 
        &=i\epsilon_{V,\nu}^{*} \bigg[ g^{\mu\nu}(m_{B}+m_{V})A_{1}(q^2) \\
        & \quad - \frac{(p+p^{\prime})^{\mu}q^{\nu}}{m_{B}+m_{V}}A_{2}(q^2) 
        - q^{\mu}q^{\nu} \frac{2m_{V}}{q^{2}} (A_{3}(q^2)-A_{0}(q^2) \bigg]\,,
\end{aligned}
    \label{eq:formfacB}
\end{align}
where $p$ and $p^{\prime}$ are the 4-momenta of the initial-state meson and final-state meson, respectively. $q^\mu = p^\mu - p^{\prime \mu} $ is the momentum transfer, $\epsilon_V$ is the polarization vector of the vector meson $V$. $m_B$, $m_P$ and $m_V$ are the masses of $B$, $P$ and $V$ respectively. The form factors $f_{0,+}$, $V_0$, and $A_{0,1,2,3}$, which depend on $q^2$, are taken from the numerical results computed using the light-cone sum rules (LCSR) approach~\cite{Ball:2004ye,Bharucha:2015bzk}.

With all necessary inputs at hand, we can write down the decay amplitudes for $B(p) \to P(p^{\prime}) \chi( k_1) \bar{\chi}(k_2)$ and $B(p) \to V(p^{\prime}) \chi( k_1) \bar{\chi}(k_2)$ as
\begin{align}
    \begin{aligned}
    i\mathcal{M}_{B\to P\chi\bar{\chi}} &=  
        C_{d\chi}^{V V}\langle P(p^{\prime})|{\bar d_i}\gamma^\mu b|B(p) \rangle \bar{u}(k_1)\gamma_\mu v(k_2)
        + C_{d\chi}^{V A}\langle P(p^{\prime})|{\bar d_i}\gamma^\mu b|B(p) \rangle \bar{u}(k_1)\gamma_\mu\gamma_5 v(k_2)\,, 
        \\[2mm]
    i\mathcal{M}_{B\to V\chi\bar{\chi}} 
        &= C_{d\chi}^{V V}\langle V(p^{\prime})|{\bar d_i}\gamma^\mu b|B(p) \rangle \bar{u}(k_1)\gamma_\mu v(k_2)
        + C_{d\chi}^{V A}\langle V(p^{\prime})|{\bar d_i}\gamma^\mu b|B(p) \rangle \bar{u}(k_1)\gamma_\mu\gamma_5 v(k_2) \\
        &\quad + C_{d\chi}^{A V} \langle V(p^{\prime})|{\bar d_i}\gamma^\mu\gamma^5 b|B(p) \rangle  \bar{u}(k_1)\gamma_\mu v(k_2) \\
        &\quad + C_{d\chi}^{A A}\langle V(p^{\prime})|{\bar d_i}\gamma^\mu\gamma^5 b|B(p)\rangle \bar{u}(k_1)\gamma_\mu\gamma_5 v(k_2) \,, 
    \end{aligned} 
\end{align}
where $d_i = d, s$ for $\pi (\rho), K (K^*)$, respectively. The corresponding differential decay widths can then be obtained through a straightforward calculation by using \texttt{FeynCalc}~\cite{Shtabovenko:2024aum}, yielding
\begin{align}
    \frac{d\Gamma_{B\to P\chi\bar{\chi}}}{dq^{2}}
       &=\frac{\lambda^{\frac{1}{2}}(m_{B}^{2},m_{P}^{2},q^2)\kappa^{\frac{1}{2}}(m_{\chi}^{2},q^2)}{384\pi^{3}m_{B}^{3}}\left\{\frac{2(q^2+2m_{\chi}^{2})\lambda(m_{B}^{2},m_{P}^{2},q^2)}{q^2}f_{+}^{2}\left|C_{d\chi}^{VV}\right|^{2}\right.\notag\\
       &+\left.\frac{2}{q^2}\left[6m_{\chi}^{2}(m_{B}^{2}-m_{P}^{2})^{2}f_{0}^{2}+(q^2-4m_{\chi}^{2})\lambda(m_{B}^{2},m_{P}^{2},q^2)f_{+}^{2}\right]\left|C_{d\chi}^{VA}\right|^{2}\right\}\,,\\
    \frac{d\Gamma_{B\to V \chi\bar{\chi}}}{dq^2}
       &=\frac{\lambda^{\frac{3}{2}}(m_B^2,m_V^2,q^2)\kappa^{\frac{1}{2}}(m_{\chi}^2,q^2)}{96\pi^3m_B^3(m_B+m_V)^2}V_0^2\left[(q^2+2m_{\chi}^2)\left|C_{d\chi}^{VV}\right|^2+(q^2-4m_{\chi}^2)\left|C_{d\chi}^{VA}\right|^2\right]\notag\\
       &+\frac{\lambda^{\frac{1}{2}}(m_B^2,m_V^2,q^2)\kappa^{\frac{1}{2}}(m_{\chi}^2,q^2)}{96\pi^3m_B^3q^2}\left\{(q^2-4m_{\chi}^2)\left[32m_{B}^{2}m_{V}^{2}A_{12}^2\right.\right.\notag\\
       &+\left.\left.(m_B+m_V)^2q^2A_1^2\right]
       +3m_{\chi}^2\lambda(m_{B}^{2},m_{V}^{2},q^2)A_0^2\right\}\left|C_{d\chi}^{AA}\right|^2\notag\\
       &+\frac{\lambda^{\frac{1}{2}}(m_B^2,m_V^2,q^2)\kappa^{\frac{1}{2}}(m_{\chi}^2,q^2)}{96\pi^3m_B^3q^2}\left\{(q^2+2m_{\chi}^2)\left[32m_{B}^{2}m_{V}^{2}A_{12}^2\right.\right.\notag\\
       &+\left.\left.(m_B+m_V)^2q^2A_1^2\right]\right\}\left|C_{d\chi1}^{AV}\right|^2\,,
\end{align}
where $q^2$ denotes the invariant mass squared of the DM pair and $m_{\chi}$ refers to the mass of the DM particle. $\kappa(m_{\chi}^2,q^2)= 1-\frac{4m_{\chi}^2}{q^2}$ and the Källen function $\lambda(x,y,z)\equiv x^2+y^2+z^2-2(xy+yz+zx)$.

\begin{figure}[t]
  \centering
  \begin{tabular}{cc}  %
    \includegraphics[width=0.48\textwidth]{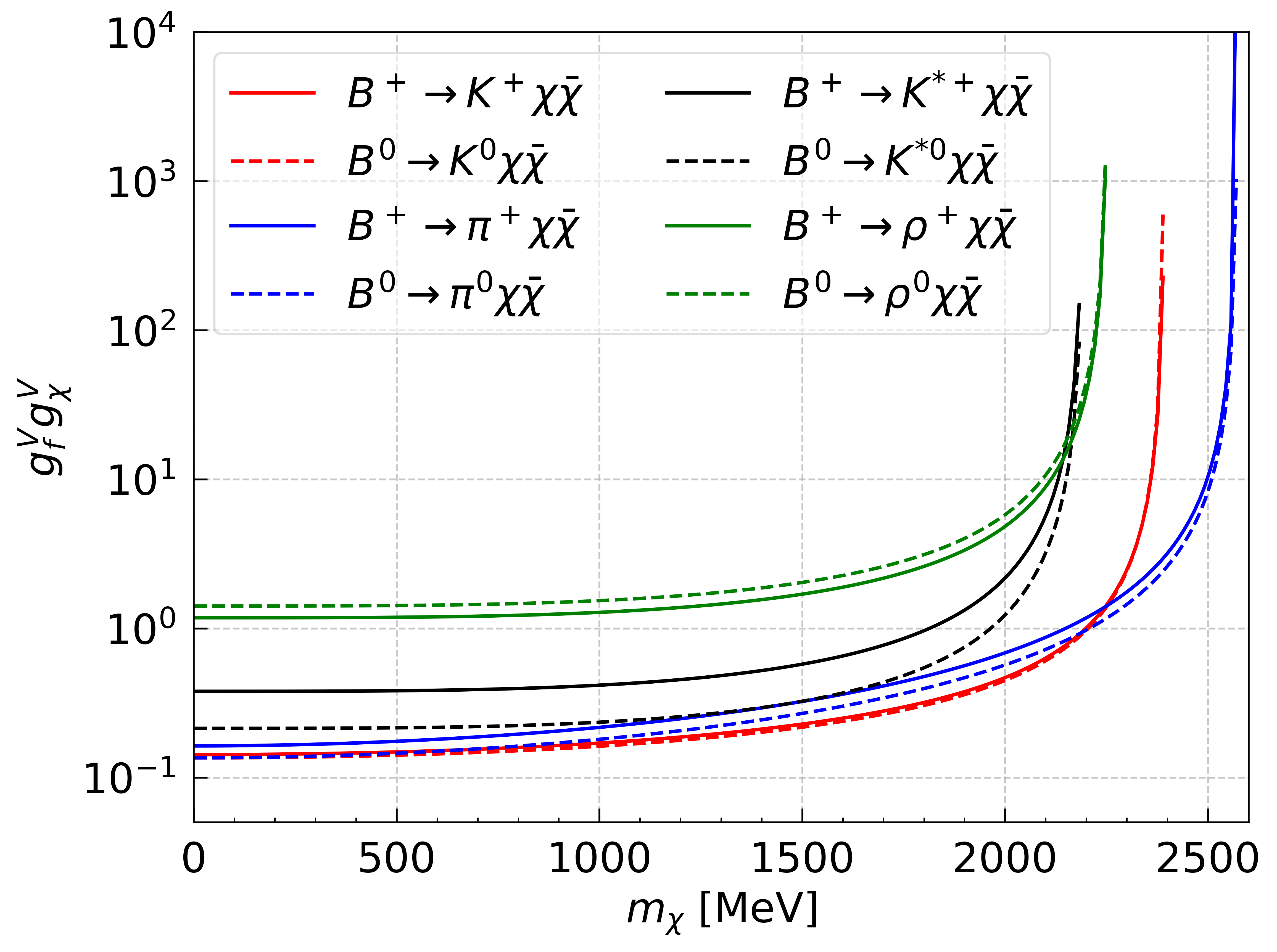} & 
    \includegraphics[width=0.48\textwidth]{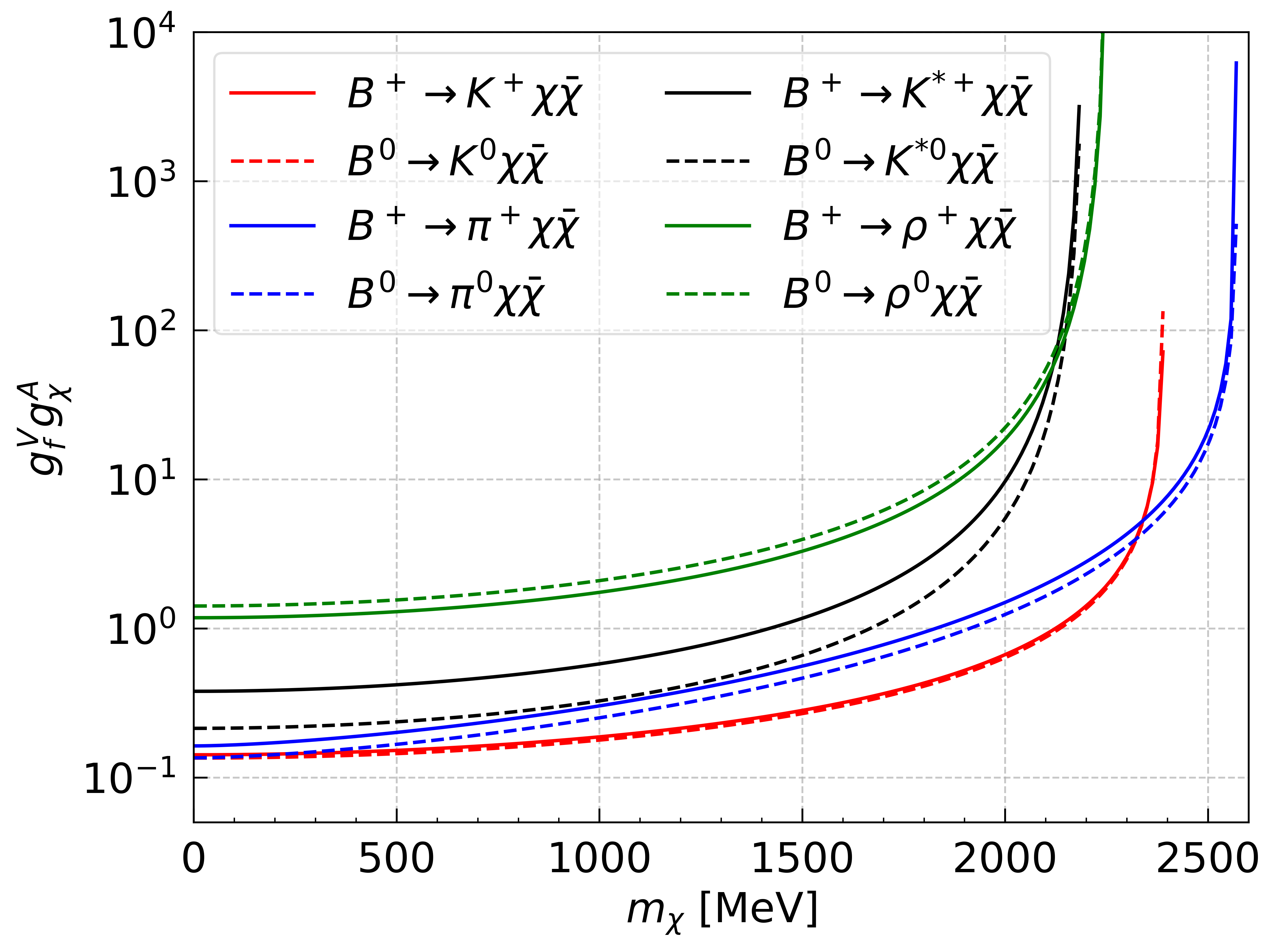} \\
    \includegraphics[width=0.48\textwidth]{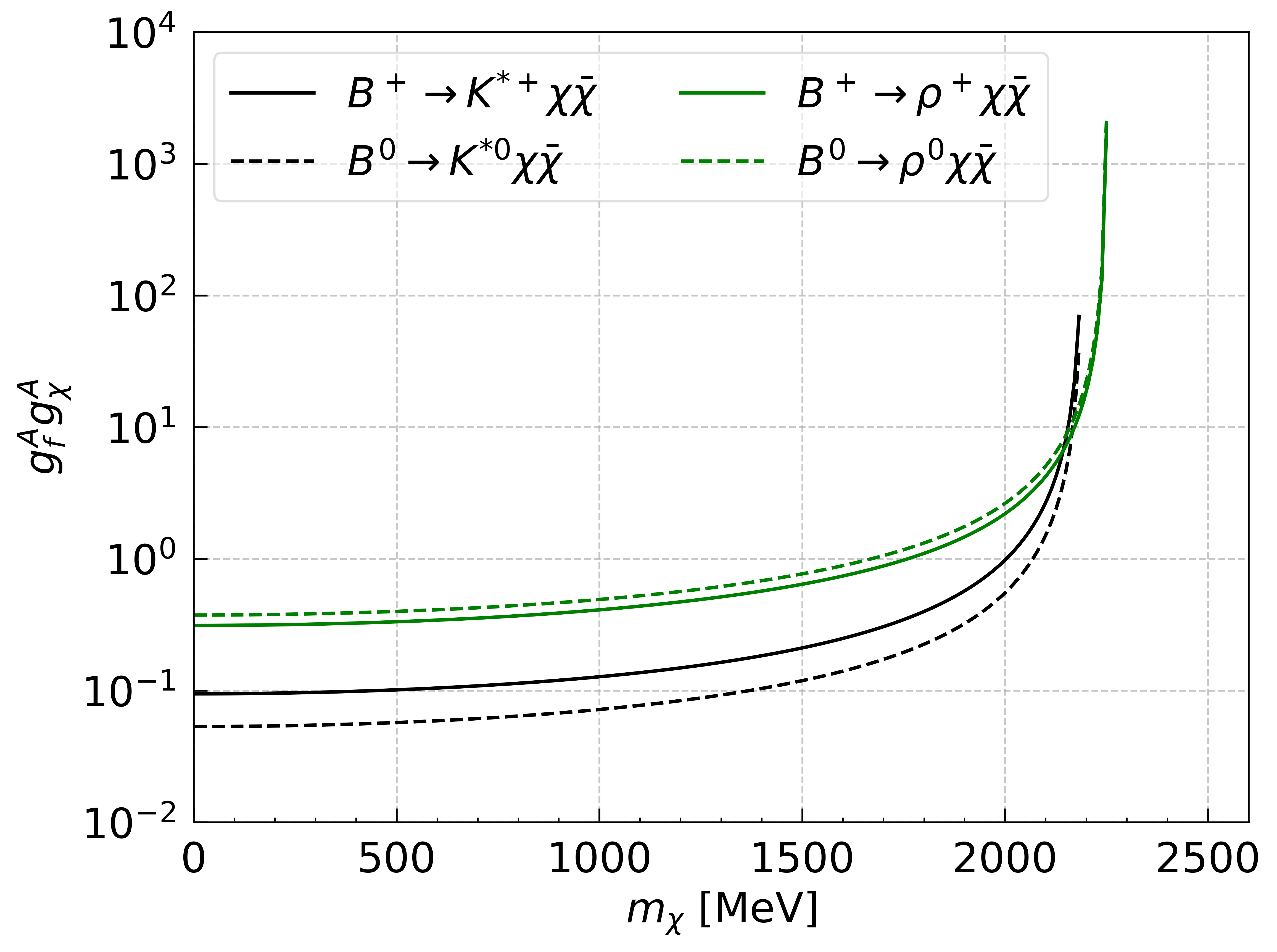} & 
    \includegraphics[width=0.48\textwidth]{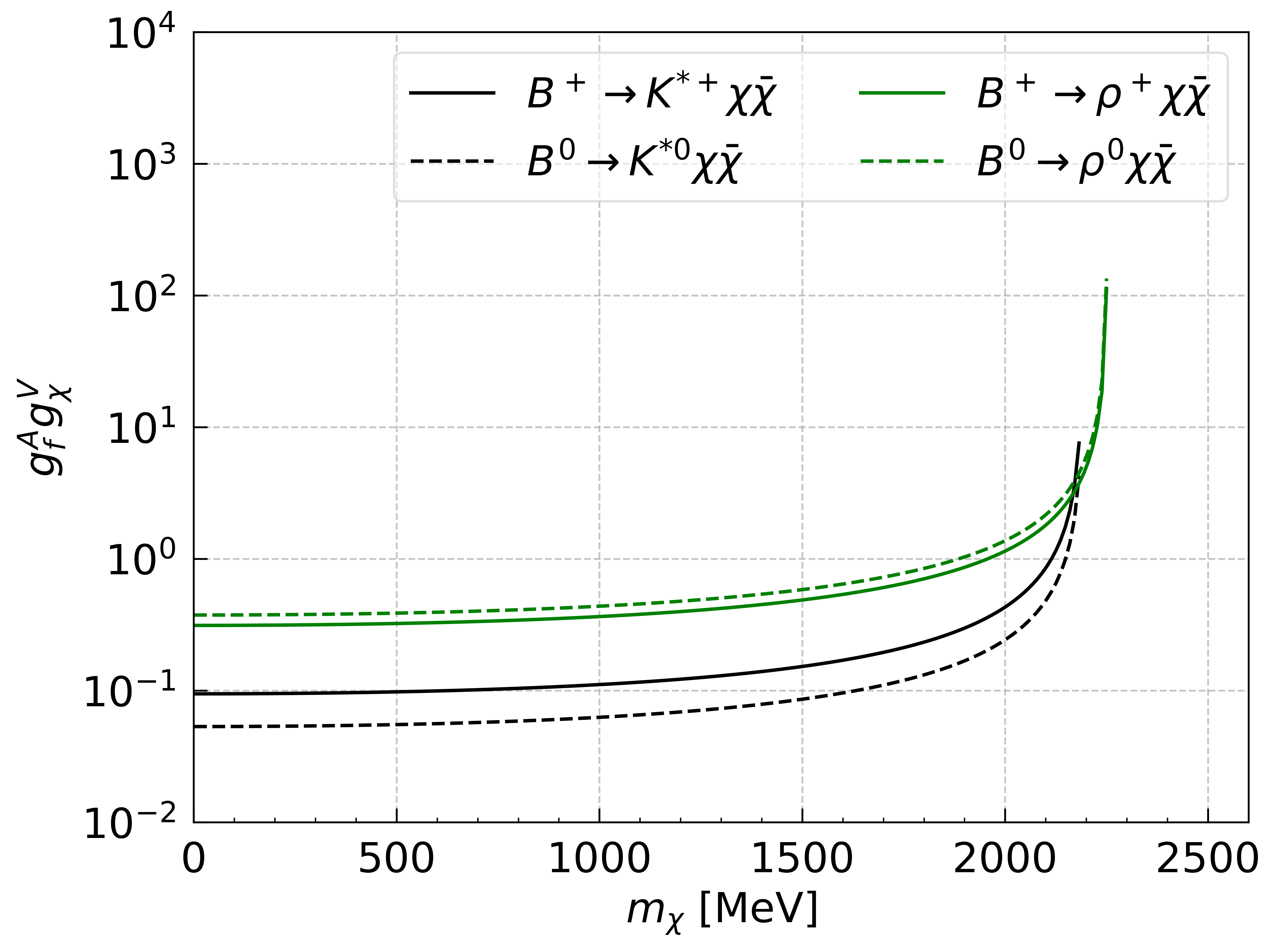} \\
  \end{tabular}
  \caption{Constraints on the effective couplings $g_f^{V(A)} g_{\chi}^{V(A)}$ versus DM mass $m_\chi$ from rare $B$ meson decays. The colored curves represent upper limits from different $b\to s(d)\chi\bar{\chi}$ transition channels, with the regions above each curve experimentally excluded at $90\%$ confidence level. }
  \label{fig:Bs}
\end{figure}

Finally, taking into account the matching conditions given in Eq.~\eqref{eq:matching}, the parameters of the $Z^\prime$ portal model can be directly constrained by the experimental measurements listed in Table~\ref{tab:Margin_settings}, as illustrated in Figure~\ref{fig:Bs}. For the coupling product $g_f^{V} g_{\chi}^{V}$, the dominant constraint arises from the $B^0\to K^0 \chi \bar{\chi}$, which is induced by the $b\to s$ transition.As shown in the figure, the constraints from the $B^+\to K^+ \chi \bar{\chi}$ and $B^0\to K^0 \chi \bar{\chi}$ channel are in good agreement. This consistency implies that, with higher experimental precision in the future, the neutral and charged processes will enable a valuable mutual cross-validation. As for the $b\to d$ transition, the most stringent constraint comes from the $B^0\to \pi^0\chi \bar{\chi}$ channel. In addition, as the DM mass $m_\chi$ approaches the kinematic endpoint $(m_B - m_{P,V})/2$, phase-space suppression rapidly degrades the constraints from a given decay channel. Hence, for $m_\chi>2.3~\mathrm{GeV}$, $B^0\to \pi^0 \chi \bar{\chi}$ will provide a stronger constraint than $B^0\to K^0 \chi\bar{\chi}$ channel due to the lighter final-state meson. For the coupling product $g_f^{V} g_{\chi}^{A}$, the constraints exhibit a similar trend as those for $g_f^V g_\chi^V$, with the main difference being an overall upward shift in the upper limits on the couplings. Since the axial-vector quark current yields no $B\to P$ hadron matrix elements, the constraints on $g_f^A g_\chi^A$ and $g_f^A g_\chi^V$ come exclusively from $B\to V$ modes, dominated by $B^0\to K^{*0} \chi\bar{\chi}$ decay channel.

\subsection{$K$ Meson Decay}
\label{sec:kaondecay}
In this subsection, we concentrate on the study of rare $K$ meson decays. Following the same approach as used for $B$-meson decays, we analyze their differential decay widths and derive corresponding constraints on the parameter space of the $Z^{\prime}$ portal model. Note that the Wilson coefficients induced by the $Z^\prime$ are almost purely real\footnote{The imaginary part of the CKM factors is ignored in the present calculation.}, resulting in no contribution to the CP-violating $K_L \to \pi^0 \chi\bar{\chi}$ process, whose amplitude is proportional to the imaginary part of the corresponding Wilson coefficients~\cite{He:2022ljo}. Therefore, we focus on the $K^+\to\pi^+ \chi\bar{\chi}$ decay channel. The hadronic matrix element for $K^+\to\pi^+$ transitions is dominated by the vector quark current. In the present work, we adopt the parameterization of this hadronic matrix element as established in Refs.~\cite{Kamenik:2011vy, He:2022ljo}, given by
\begin{align}
   \begin{aligned}
      \langle  \pi^+(p^{\prime}) | {\bar s} \gamma^\mu d | K^+(p) \rangle \simeq \left( p+p^{\prime} \right)^\mu\,.
      \label{eq:formfacK2pi}
   \end{aligned}    
\end{align}

The amplitude for the $K^+(p) \to \pi^+(p^{\prime}) \chi(k_1) \bar{\chi}(k_2)$ process can be written directly as
\begin{align}
   \begin{aligned}
      i\mathcal{M}_{K^+\to \pi^+\chi\bar{\chi}}
       &=  C_{d\chi}^{V V}\langle \pi^+(p^{\prime})|{\bar s}\gamma^\mu d|K^+(p) \rangle
      \bar{u}(k_1)\gamma_\mu v(k_2)\\
      &+ C_{d\chi}^{V A}\langle \pi^+(p^{\prime})|{\bar s}\gamma^\mu d|K^+(p) \rangle \bar{u}(k_1)\gamma_\mu\gamma_5 v(k_2)\,. 
   \end{aligned} 
\end{align}
Consequently, the differential decay width is also computed using \texttt{FeynCalc}, with the explicit expression given below:
\begin{align}
   \begin{aligned}
       \frac{d\Gamma_{K^+\to \pi^+\chi\bar{\chi}}}{dq^{2}}
        &=\frac{\lambda^{\frac{1}{2}}(m_{K}^{2},m_{\pi}^{2},q^2)\kappa^{\frac{1}{2}}(m_{\chi}^{2},q^2)}{384\pi^{3}m_{K}^{3}}\left\{\frac{2(q^2+2m_{\chi}^{2})\lambda(m_{K}^{2},m_{\pi}^{2},q^2)}{q^2}\left|C_{d\chi}^{VV}\right|^{2}\right.\\
        &+\left.\frac{2}{q^2}\left[6m_{\chi}^{2}(m_{K}^{2}-m_{\pi}^{2})^{2}+(q^2-4m_{\chi}^{2})\lambda(m_{K}^{2},m_{\pi}^{2},q^2)\right]\left|C_{d\chi}^{VA}\right|^{2}\right\}\,.
    \label{eq:Ktopidecay}
   \end{aligned}    
\end{align}

\begin{figure}[t]
  \centering
  \begin{tabular}{cc}  %
    \includegraphics[width=0.48\textwidth]{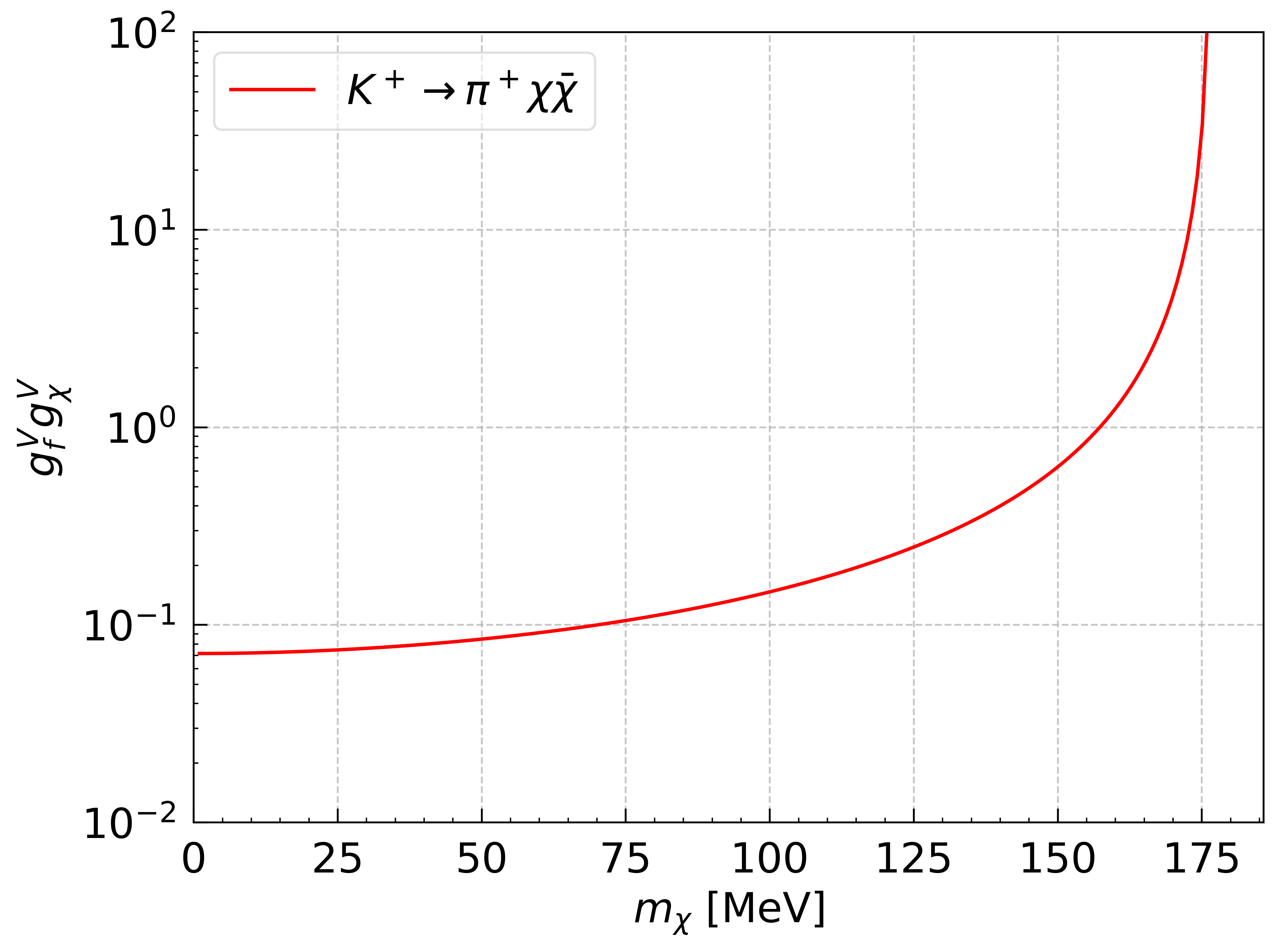} & 
    \includegraphics[width=0.48\textwidth]{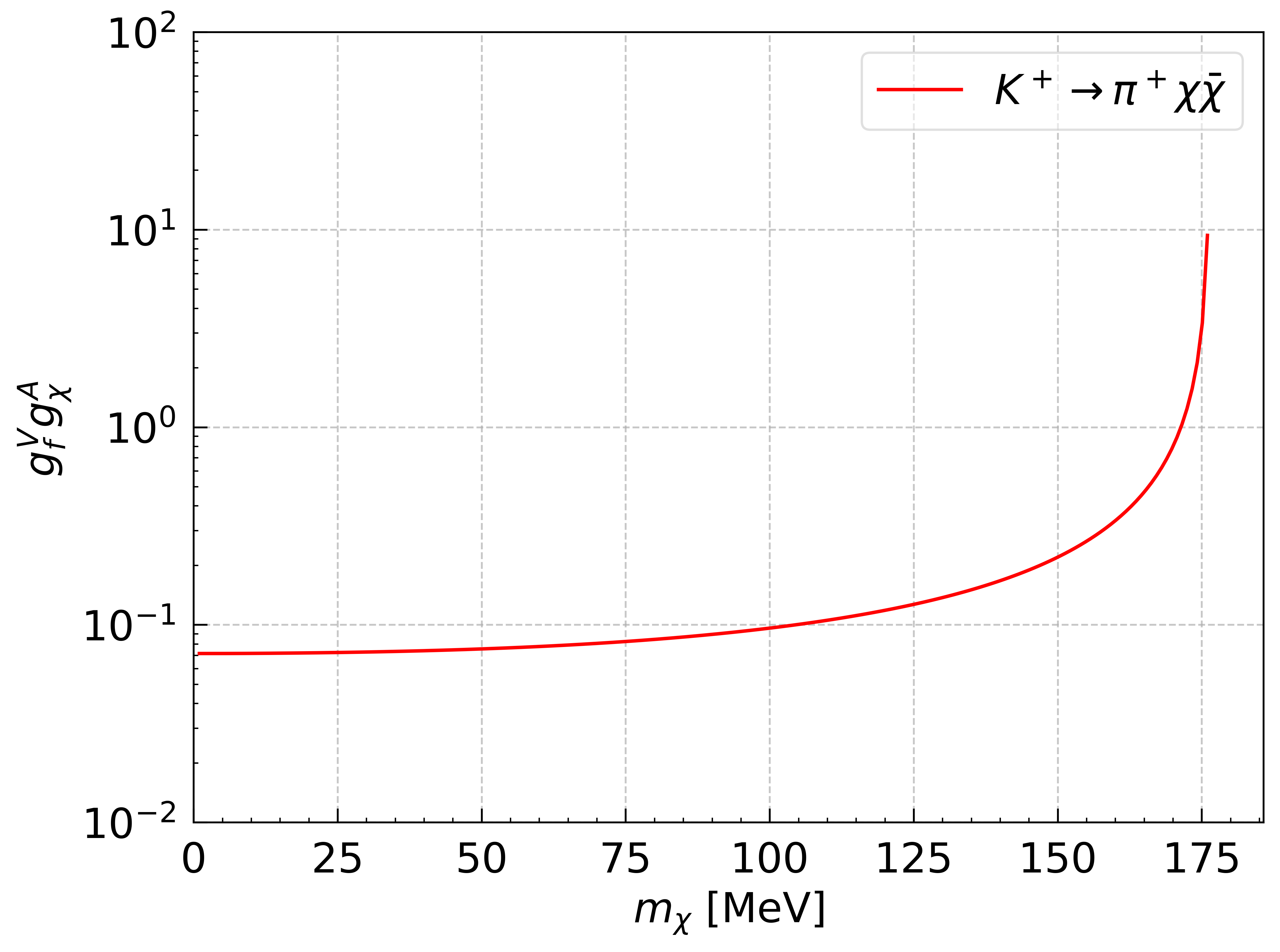} \\
  \end{tabular}
\caption{Same as Figure~\ref{fig:Bs}, but for rare $K$ meson decays.}
  \label{fig:Ks}
\end{figure}

Herein, we elaborate on the constraints on the parameter space of the $Z^{\prime}$ portal model imposed by $K$ meson decays. As shown in Figure~\ref{fig:Ks}, the exclusion limits for the coupling product $g_f^{V} g_{\chi}^{V(A)}$ are found to be on the order of $10^{-1}$. Notably, the constraint associated with $g_f^{V} g_{\chi}^{A}$ is marginally more stringent, and this disparity stems from the distinct decay widths of the relevant $K$ meson channels.

\section{Complementarity of rare meson decays with DM direct detection}
\label{sec:complementarity}

DM direct detection experiments have been primarily dedicated to searching for WIMPs, with masses from GeV to TeV scale. However, their sensitivity to light DM at the GeV scale and below is significantly weaker. This is because the kinetic energy of light DM particles is small and very little gets transferred to a nucleon in elastic scattering. In this section, we use the upper bounds on the $Z^{\prime}$ portal model parameters derived from rare meson decays in the previous section to constrain the scattering cross section of light dark matter off nucleons.

\begin{table}[tb]
  \centering
  \begin{tblr}{
  colspec = {Q[2cm,c] Q[4cm,c] Q[3cm,c] Q[5.5cm,c]},
  rows = {abovesep=4pt, belowsep=4pt},
  width = 0.95\textwidth,
  column{1} = {leftsep=2pt, rightsep=2pt},
}
    \hline
    & Operator & Structure & DM-nucleon Cross Section \\ \hline
    $\mathcal{O}_{d\chi}^{V V}$  & $(\bar{d}_i\gamma^\mu d_j)(\bar{\chi}\gamma_\mu\chi)$ & SI, MI & $\sim A^2$ \\ \hline
    $\mathcal{O}_{d\chi}^{V A}$  & $(\bar{d}_i\gamma^\mu d_j)(\bar{\chi}\gamma_\mu\gamma_5\chi)$ & SI, MD & $\sim v^2$ \\ \hline
    $\mathcal{O}_{d\chi}^{A V}$  & $(\bar{d}_i \gamma^\mu\gamma_5 d_j)(\bar{\chi}\gamma_\mu\chi)$ & SD, MD & $\sim v^2$ \\ \hline
    $\mathcal{O}_{d\chi}^{A A}$  & $(\bar{d}_i \gamma^\mu\gamma_5 d_j)(\bar{\chi}\gamma_\mu\gamma_5\chi)$ & SD, MI & $\sim S^2_A(p)$ \\ \hline
    \end{tblr}
  \caption{The properties of operators induced by the $Z^{\prime}$ portal model for DM direct detection~\cite{An:2012va}. The table shows whether the DM-nucleon cross section is independent (SI) or dependent (SD) on the nucleon spin, and whether it is dependent on the DM momentum (MD) or not (MI). $A$ denotes the target nucleon number and $S_A(p)$ is the axial-vector structure factor~\cite{Klos:2013rwa}.}
\label{operators2}
\end{table}

DM direct detection experiments are broadly categorized into spin-independent (SI) and spin-dependent (SD) searches. For SI interactions, the scattering cross section scales with the square of the target nucleon number $A$, which motivates the use of heavy nuclei to enhance the signal. Therefore, SI searches generally place stronger constraints on the cross section than SD searches. In the $Z^{\prime}$ portal model, the induced operators involving a vector quark current, i.e. $\mathcal{O}_{d\chi}^{V V}$ and $\mathcal{O}_{d\chi}^{V A}$, correspond to spin-independent interactions, while the operators with an axial-vector quark current, i.e. $\mathcal{O}_{d\chi}^{A V}$ and $\mathcal{O}_{d\chi}^{A A}$, give rise to spin-dependent interactions. The DM-nucleon cross section induced by operators $\mathcal{O}_{d\chi}^{V A}$ and $\mathcal{O}_{d\chi}^{A V}$ depend on the momentum exchange~\cite{An:2012va}. Given the typical velocity of the DM in the galactic halo $v\sim 10^{-3}$, these cross sections are highly suppressed by a factor of $v^2$. The properties of these operators are collected in Table~\ref{operators2}. Next, we focus on analyzing the constraints on the DM-nucleon cross section associated with operators $\mathcal{O}_{d\chi}^{V V}$ and $\mathcal{O}_{d\chi}^{A A}$. Using the joint constraints on the DM mass and coupling products from rare meson decays, both SI and SD DM-nucleon cross sections can be calculated by package \texttt{micrOMEGAs}~\cite{Belanger:2020gnr}.

\begin{figure}[tb]
  \centering
  \begin{tabular}{cc}  %
    \includegraphics[width=0.48\textwidth]{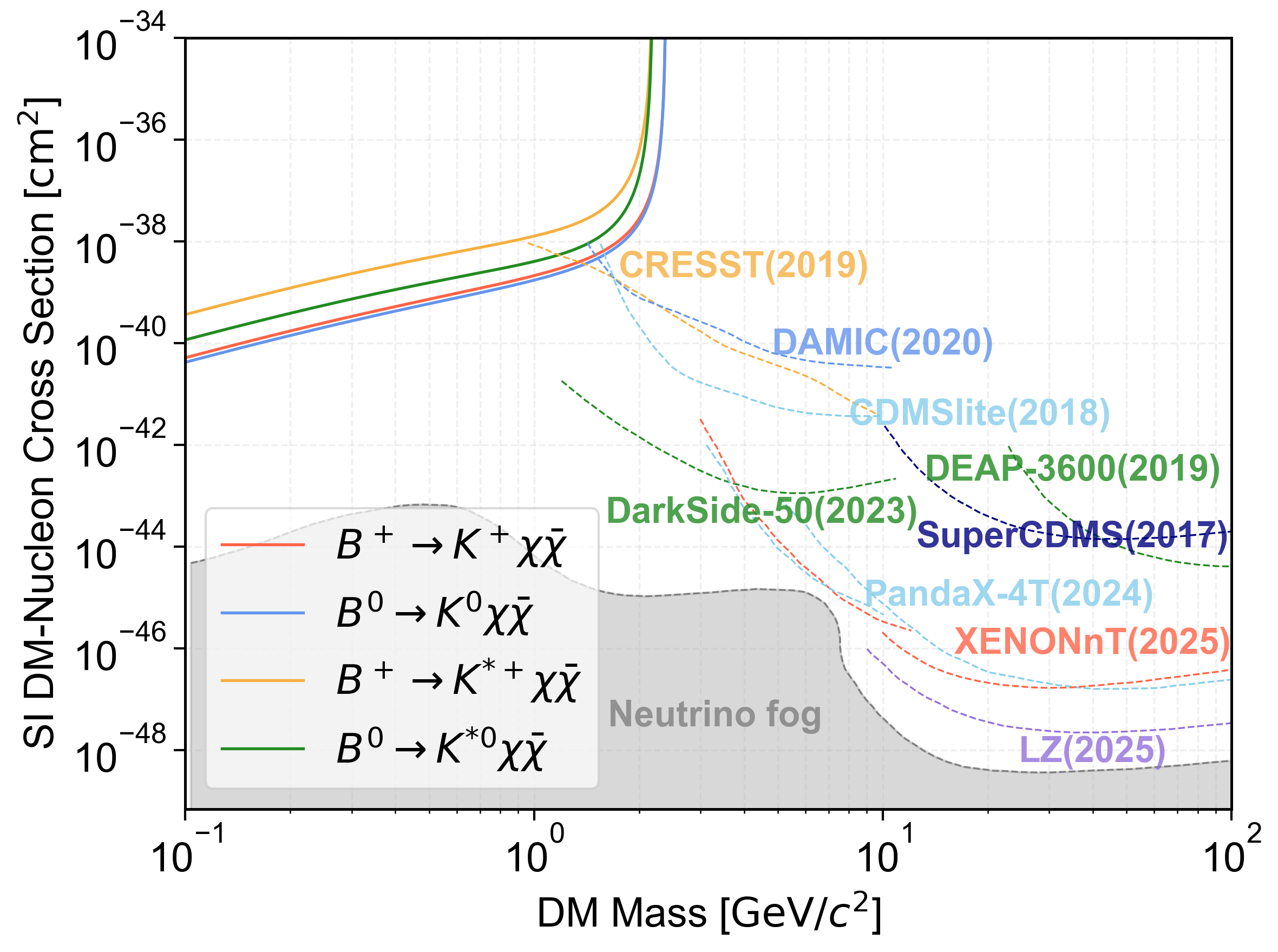}
     \includegraphics[width=0.48\textwidth]{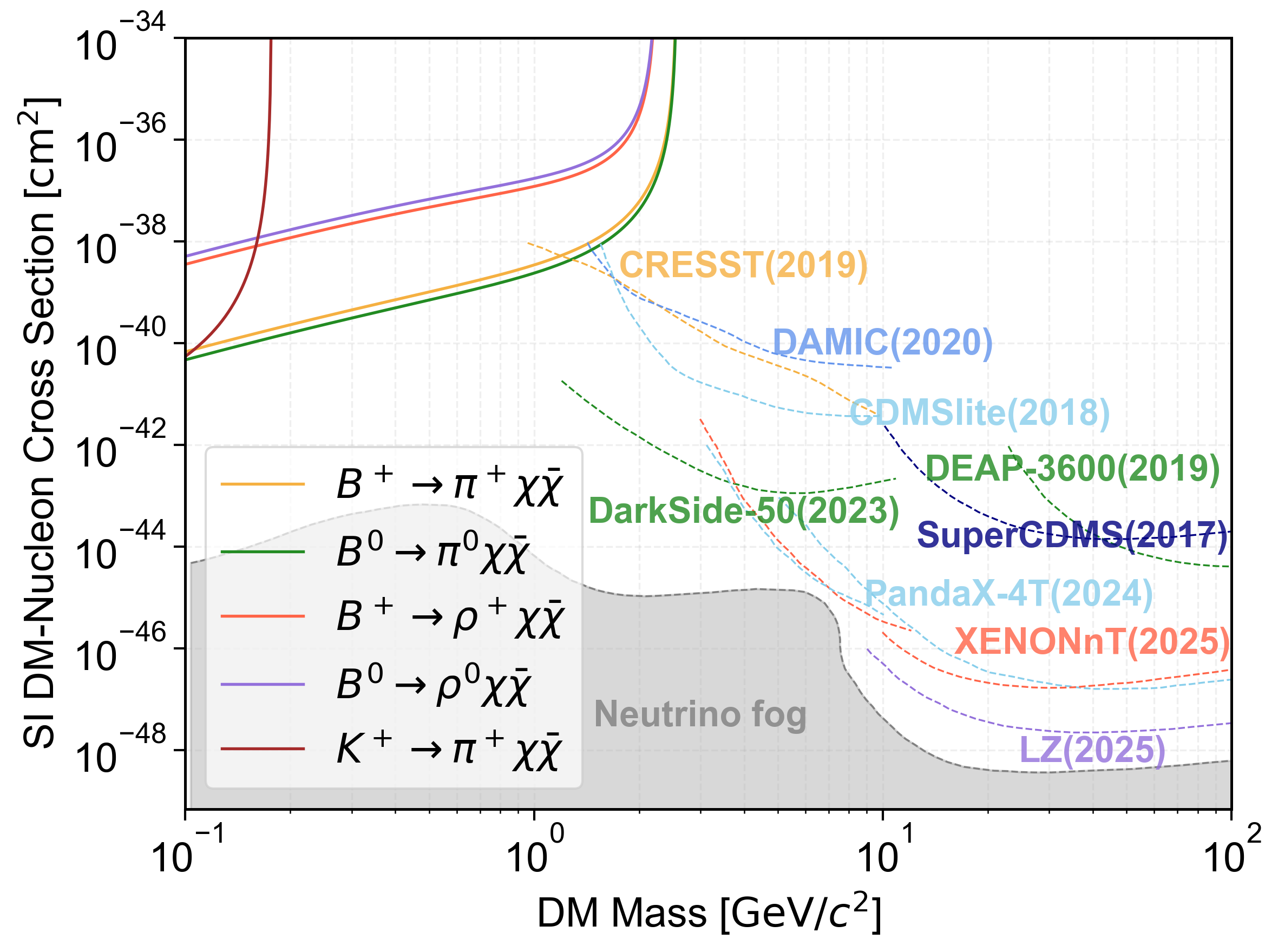}
  \end{tabular}
  \caption{Upper limits on spin-independent DM-nucleon cross section as a function of the DM mass. Also shown are the CRESST(2019)~\cite{CRESST:2019jnq}, DAMIC(2020)~\cite{DAMIC:2020cut},  CDMSlite(2018)~\cite{SuperCDMS:2018gro},  DarkSide-50(2023)~\cite{DarkSide-50:2022qzh},  SuperCDMS(2017)~\cite{SuperCDMS:2017mbc},                   DEAP-3600(2019)~\cite{DEAP:2019yzn}, LZ(2025)~\cite{LZ:2024zvo}, PandaX-4T(2024)~\cite{PandaX:2024qfu} and XENONnT(2025)~\cite{XENON:2025vwd} limits.}
  \label{fig:SI-DM}
\end{figure}

In Figure~\ref{fig:SI-DM}, we compare the constraints on the SI DM-nucleon cross section from rare $B$ and $K$ meson decays with those from various DM direct detection experiments. To avoid overlapping curves, the results are presented in two separate panels. Here, we take DM mass as the independent variable and fix $Z^{\prime}$ mass $m_{Z^{\prime}}=100\, \text{GeV}$. As shown in the figure, DarkSide-50 (2023) provides the strongest direct-detection limit in the few-GeV region, while rare meson decays offer uniquely sensitive constraints on DM for masses below $1~\mathrm{GeV}$. Among these, the limit from $B^0 \to K^0 \chi \bar{\chi}$ is the most stringent, constraining the DM-nucleon scattering cross section to below $10^{-40} ~\mathrm{cm}^2$. It is well known that as direct detection experiments approach the neutrino floor, distinguishing DM signals from the coherent neutrino background becomes difficult. In contrast, constraints derived from rare meson decays remain unaffected by this background. With improved experimental precision in these decays, the resulting limits on the parameter space of light DM will be pushed below the neutrino floor, providing a complementary and background-free probe. As shown in Figure~\ref{fig:Bs}, the constraints from rare meson decays are comparable for both $g_f^Vg_{\chi}^V$ and $g_f^Vg_{\chi}^A$ interactions. Therefore, the solid lines in Figure~\ref{fig:SI-DM} can roughly be extended to represent the SI momentum-dependent interaction scenario.

\begin{figure}[tb]
  \centering
  \begin{tabular}{cc}  %
    \includegraphics[width=0.48\textwidth]{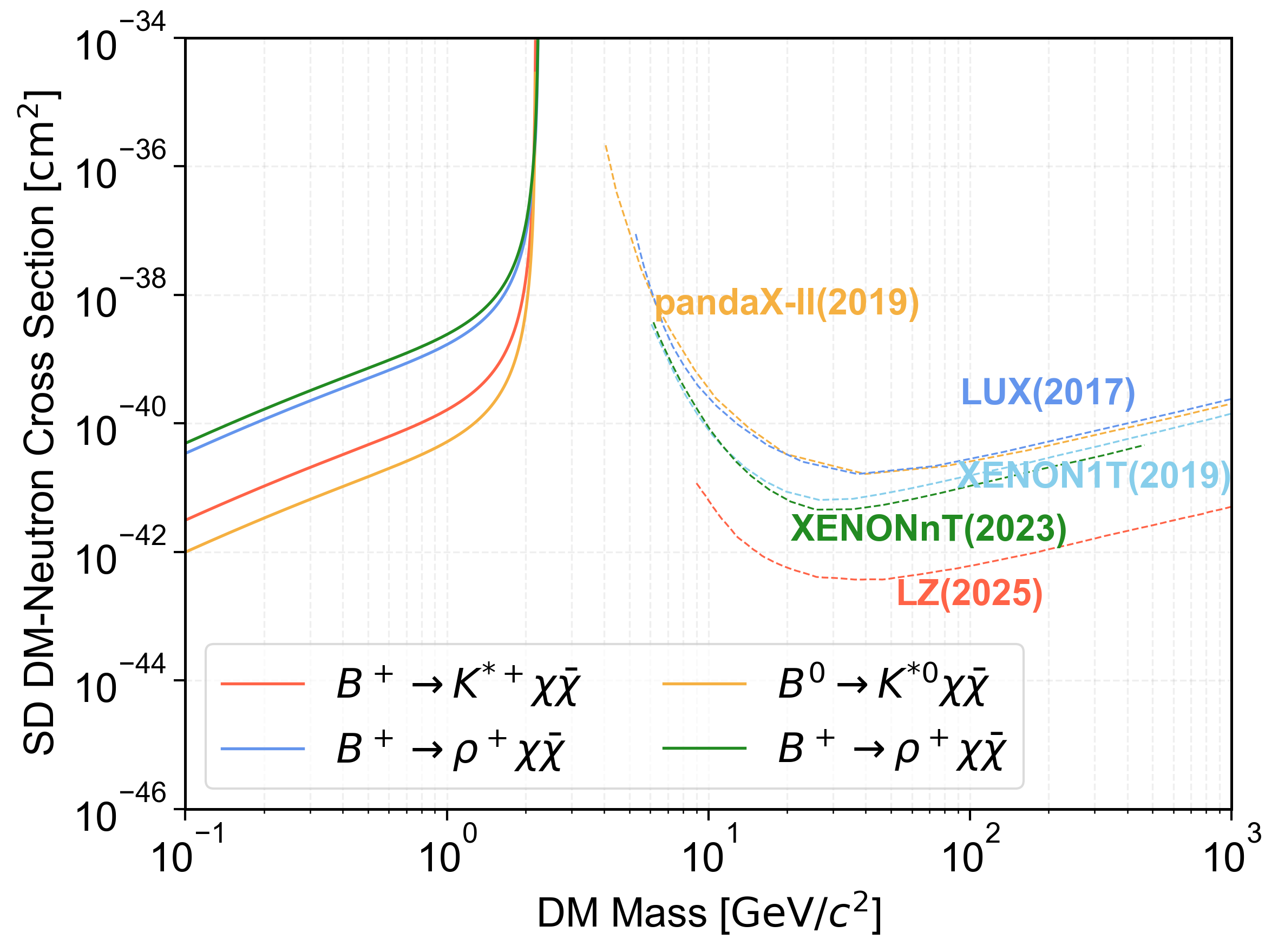}
    \includegraphics[width=0.48\textwidth]{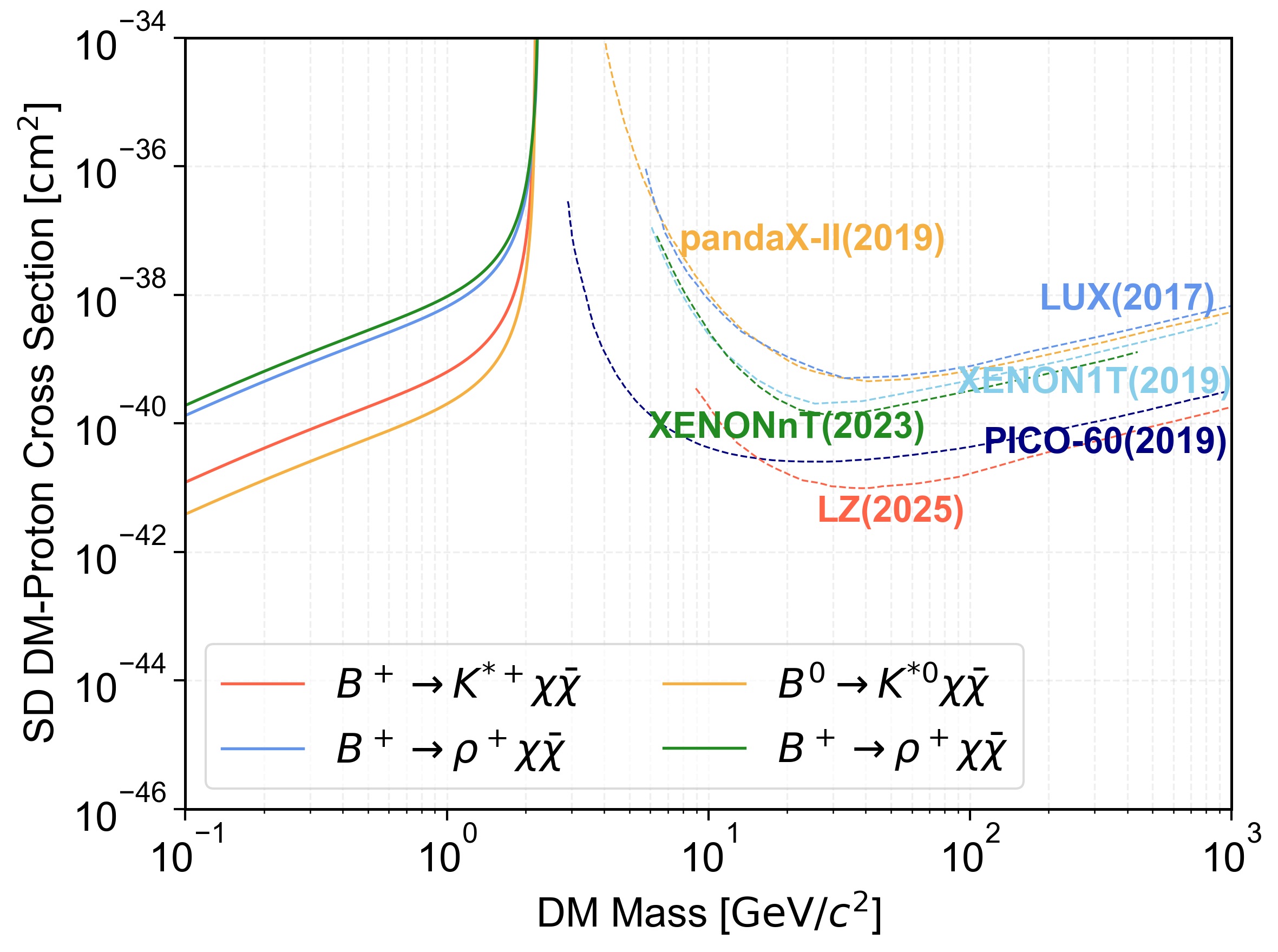}
  \end{tabular}
  \caption{Upper limits on spin-dependent DM-neutron(left) and DM-proton(right) cross sections as a function of the DM mass. Also shown are the PICO-60~\cite{PICO:2019vsc}, PandaX-II~\cite{PandaX-II:2018woa}, LUX~\cite{LUX:2017ree}, XENONnT~\cite{XENON:2023cxc}, XENON1T~\cite{XENON:2019rxp} and LZ~\cite{LZ:2024zvo} limits.} 
  \label{fig:SD-DM}
\end{figure}

Figure~\ref{fig:SD-DM} shows the constraints on the spin-dependent DM-neutron(proton) cross section from rare $B$ and $K$ meson decays. For comparison, constraints from the LUX, XENON1T, XENONnT, PICO-60, and LZ experiments on the spin-dependent DM-neutron(proton) cross section are also shown in the figure. Among these experiments, the LZ experiment provides the strongest constraints for DM masses above 10 GeV for both spin-dependent DM-neutron and DM-proton cross sections. For DM mass in the few-GeV region, the PICO-60 experiment places the strongest constraints for spin-dependent DM-proton cross section. The solid lines in Figure~\ref{fig:SD-DM} provide constraints on the spin-dependent DM-neutron(proton) cross section derived from the rare meson decays. Among these, the most stringent limit is placed by $B^0 \to K^{*0} \chi\bar{\chi}$, which is on the order of $10^{-42}\, \text{cm}^2$ to $10^{-40}\, \text{cm}^2$ for light DM in the mass range 100 MeV to 1 GeV. The solid lines in Figure~\ref{fig:SD-DM} can also roughly be extended to represent the SD momentum-dependent interaction scenario.

\section{Summary and conclusions}
\label{sec:conclusion}
In this work, we explore the possibility of probing light DM via the rare decays of mesons, and discuss the complementarity between the rare meson decays and DM direct detection. This work focuses on a generic $Z^{\prime}$ portal model, where a heavy neutral $Z^{\prime}$ serves as the portal between Dirac DM and SM particles. We use the DLEFT to analyze the effective operators contributing to $B \to (K^{(*)}, \pi, \rho) + \chi\bar{\chi}$ and $K \to \pi + \chi\bar{\chi}$, and calculate the upper limits of the parameters of the $Z^{\prime}$ portal model imposed by the rare decays of $B$ and $K$ mesons. To explore the complementarity between rare $B$ and $K$ meson decays and DM direct detection, we utilize these upper limits to compute the corresponding spin-dependent and spin-independent scattering cross sections between DM and nucleons.

The results indicate that $B$ and $K$ meson decays with missing energy in the final state can be efficient probe of light DM, one that is free from the neutrino floor background and thus complementary to direct detection. In the spin-independent case, the constraints from rare meson decays on the cross section of sub-GeV DM reach sensitivities comparable to those of direct detection experiments in the few-GeV region. For the spin-dependent case, these decays provide constraints that are stronger than the most stringent limits from current direct detection experiments. Furthermore, DM direct detection experiments suffer from DM velocity suppression and are therefore insensitive to the scattering cross sections associated with the operators $\mathcal{O}_{d\chi}^{V A}$ and $\mathcal{O}_{d\chi}^{A V}$. In contrast, rare meson decays are free of this suppression and can impose constraints of similar magnitude to those from the operators $\mathcal{O}_{d\chi}^{V V}$ and $\mathcal{O}_{d\chi}^{A A}$. The limits derived in this paper have important implications for dark matte model building and light WIMPs searching. Given the strong motivation WIMPs and the challenges faced by direct detection at sub-GeV mass, it is valuable to extend this work to other light DM models.

\section*{Acknowledgments}
The work is supported by the National Key R\&D Program of China under Grant No. 2023YFA1606000, the Science and Technology R\&D Program Joint Fund Project of Henan Province under Grant No. 225200810030, the Natural Science Foundation of Henan Province
under Grant No. 242300421390, and the National Natural Science Foundation of China under Grant No. 12547187.

\bibliographystyle{utphys}
\bibliography{references}

\end{document}